\documentclass[12pt]{iopart}

\usepackage{amsmath,amssymb}
\usepackage{graphicx}% Include figure files
\usepackage{dcolumn}% Align table columns on decimal point
\usepackage{bm}% bold math
\usepackage{color}
\usepackage{longtable}
\usepackage{cite}

\usepackage{mathtools}

\newcommand{\nc}{\newcommand}
\nc{\bra}{\langle}
\nc{\ket}{\rangle}
\nc{\vac}{|0\ket}
\nc{\da}{^{\dagger}}
\nc{\HASEP}{\mathcal{H}_{\text{ASEP}}}
\nc{\HLK}{\mathcal{H}_{\text{ASEP-LK}}}
\nc{\THLK}{\Tilde{\mathcal{H}}_{\text{ASEP-LK}}}
\nc{\so}{\hat{S}}
\nc{\nm}{\hat{n}}
\nc{\im}{\text{i}}
\nc{\Imath}{\mathcal{I}}
\nc{\blue}{\textcolor{blue}}
\nc{\oa}{\omega_{\text{a}}}
\nc{\od}{\omega_{\text{d}}}

\nc{\tot}{\text{T}}

\begin{document}

\title[Exact analysis of the 2D ASEP with attachment and detachment of particles.]{Exact analysis of the two-dimensional asymmetric simple exclusion process with attachment and detachment of particles}

\author{Yuki Ishiguro$^{1,2}$, Jun Sato$^1$}

\address{ $^1$Faculty of Engineering, Tokyo Polytechnic University, 5-45-1 Iiyama-minami, Atsugi, Kanagawa 243-0297, Japan\\
$^2$The Insitute for Solid State Physics, The University of Tokyo, 5-1-5 Kashiwanoha, Kashiwa, Chiba 277-8581, Japan}
% \ead{y.ishiguro@eng.t-kougei.ac.jp; j.sato@eng.t-kougei.ac.jp;}
% \vspace{10pt}
% \begin{indented}
% \item[]May 2024
% \end{indented}

\begin{abstract}
The asymmetric simple exclusion process (ASEP) is a paradigmatic driven-diffusive system that describes the asymmetric diffusion of particles with hardcore interactions in a lattice. 
Although the ASEP is known as an exactly solvable model, most exact results are limited to one-dimensional systems. 
Recently, the exact steady state in the multi-dimensional ASEP has been proposed \cite{ishiguro2024exact}. The research focused on the situation where the number of particles is conserved. 
In this paper, we consider the two-dimensional ASEP with the attachment and detachment of particles (ASEP-LK), where particle number conservation is violated. By employing the result in Ref. \cite{ishiguro2024exact}, we construct the exact steady state of the ASEP-LK and reveal its properties through the exact computation of physical quantities.
\end{abstract}

%
% Uncomment for keywords
%\vspace{2pc}
%\noindent{\it Keywords}: XXXXXX, YYYYYYYY, ZZZZZZZZZ
%
% Uncomment for Submitted to journal title message
%\submitto{\JPA}
%
% Uncomment if a separate title page is required
%\maketitle
% 
% For two-column output uncomment the next line and choose [10pt] rather than [12pt] in the \documentclass declaration
%\ioptwocol
%

\section{Introduction}
Driven-diffusive systems play a central role in exploring the physics of nonequilibrium systems. A variety of complex nonequilibrium phenomena, such as biological transport\cite{macdonald1968kinetics,klumpp2003traffic} and traffic flow \cite{schadschneider2000statistical,schadschneider2010stochastic}, can be modeled by driven-diffusive systems. 
Among them, the asymmetric simple exclusion process (ASEP) is actively studied as a fundamental model for describing nonequilibrium transportation phenomena \cite{Derrida_1993,derrida1998exactly,blythe2007nonequilibrium,Crampe_2014,essler1996representations,golinelli2006asymmetric,gwa1992bethe,kim1995bethe,Golinelli_2004,Golinelli_2005,PhysRevE.85.042105,Motegi_2012,prolhac2013spectrum,prolhac2014spectrum,prolhac2016extrapolation,prolhac2017perturbative,ishiguro2023,de2005bethe,deGier_2006,deGier_2008,deGier_2011,Wen_2015,Crampe_2015,Sandow_1994,schutz1997duality,relation,schadschneider2000statistical,schadschneider2010stochastic,macdonald1968kinetics,klumpp2003traffic,bertini1997stochastic,PhysRevLett.104.230602}.
The ASEP is a stochastic process describing the asymmetric diffusion of particles with hardcore interactions in a lattice.
Although the model is simple, it contains rich nonequilibrium physics, including boundary-induced phase transitions\cite{blythe2007nonequilibrium}  and the KPZ universality class\cite{bertini1997stochastic,PhysRevLett.104.230602}. 
In addition, the ASEP is an exactly solvable model and has attracted attention in the context of mathematical physics \cite{Derrida_1993,derrida1998exactly,blythe2007nonequilibrium,Crampe_2014,essler1996representations,golinelli2006asymmetric,gwa1992bethe,kim1995bethe,Golinelli_2004,Golinelli_2005,PhysRevE.85.042105,Motegi_2012,prolhac2013spectrum,prolhac2014spectrum,prolhac2016extrapolation,prolhac2017perturbative,ishiguro2023,de2005bethe,deGier_2006,deGier_2008,deGier_2011,Wen_2015,Crampe_2015,Sandow_1994,schutz1997duality}.
In the one-dimensional (1D) case, we can evaluate the physical quantities exactly by using appropriate methods, such as the matrix product ansatz \cite{Derrida_1993,derrida1998exactly,blythe2007nonequilibrium,Crampe_2014,essler1996representations} and the Bethe ansatz \cite{golinelli2006asymmetric,gwa1992bethe,kim1995bethe,Golinelli_2004,Golinelli_2005,PhysRevE.85.042105,Motegi_2012,prolhac2013spectrum,prolhac2014spectrum,prolhac2016extrapolation,prolhac2017perturbative,ishiguro2023,de2005bethe,deGier_2006,deGier_2008,deGier_2011,Wen_2015,Crampe_2015}.
However, most of the studies on the exact analysis have been focused on 1D systems, and the exact results of systems beyond one dimension are limited to specific situations \cite{Pronina_2004,Tsekouras_2008,PhysRevE.84.061141,Ezaki_2012,Lee_1997,PhysRevE.60.6465,wang2017dynamics,wang2018analytical}.

Recently, the exact steady states of the ASEP in arbitrary dimensions have been proposed \cite{ishiguro2024exact}. 
Although the study has presented the crucial concept for constructing the solutions of exactly solvable models in more than one dimension, it focused on the situation where the number of particles is conserved.
However, many complex nonequilibrium phenomena are often modeled as open systems where particles flow into and out of the system from the environment. 
Extending the theory to the situation where particle number conservation is violated is important for understanding more diverse nonequilibrium phenomena.

\begin{figure}[bth]
    \centering
    \includegraphics[height=6.5cm]{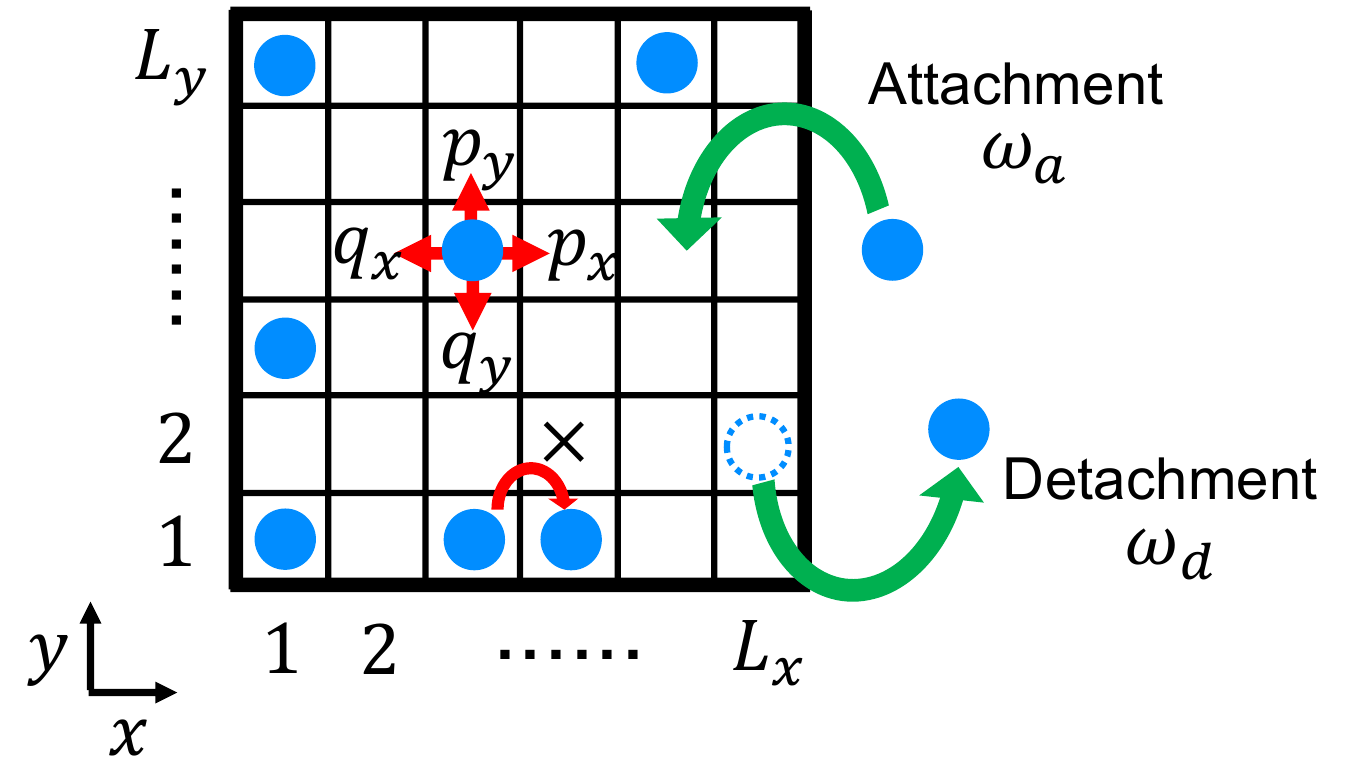}
    \caption{Asymmetiric simple exclusion process with Langmuir kinetics (ASEP-LK) in a two-dimensional lattice.}
    \label{fig:ASEP-LK_fig}
\end{figure}

Various extensions of the ASEP have been devised to describe a range of phenomena. The ASEP with Langmuir kinetics (ASEP-LK) is one of the representative extensions \cite{PhysRevLett.90.086601,PhysRevE.70.046101,PhysRevE.68.026117,Ezaki_2012_LK,PhysRevE.93.042113,PhysRevE.97.032135,PhysRevE.105.014128,DHIMAN20162038,Verma_2015}.
In the model, the ASEP is extended by introducing the attachment and detachment of particles in the bulk (Langmuir kinetics), which violates particle number conservation even in periodic and closed boundary conditions.
The schematic drawing of the 2D ASEP-LK is shown in Fig. \ref{fig:ASEP-LK_fig}.
In the 1D case, the exact steady state is constructed under the periodic boundary conditions \cite{Ezaki_2012_LK,PhysRevE.93.042113}, and that in infinitesimal Langmuir kinetics is conjectured under closed boundary conditions \cite{PhysRevE.97.032135}. 
As a study of the ASEP-LK in a 2D lattice, there are researches that extend it to multi-lane systems \cite{PhysRevE.105.014128,DHIMAN20162038,Verma_2015}.
However, in the 2D case, the ASEP-LK is mainly investigated through the mean-field approach, and exact results have not been obtained.

In this paper, we construct the exact steady states of the 2D ASEP-LK and reveal their properties based on the exact analysis of physical quantities. We focus on three types of boundary conditions: torus, closed, and multi-lane boundary conditions. In the torus (closed) boundary conditions, we consider the periodic (closed) boundary conditions both in $x$- and $y$-directions. In the multi-lane boundary conditions, we consider the closed boundary conditions in $x$-direction and the periodic boundary conditions in $y$-direction. 
By employing the exact results for the standard 2D ASEP without Langmuir kinetics \cite{ishiguro2024exact}, we construct the exact steady state of the ASEP-LK in the periodic boundary conditions and that in the infinitesimal Langmuir kinetics in the closed and multi-lane boundary conditions.
Based on the results, we clarify the effect of Langmuir kinetics and two-dimensionality on the properties of the steady state through the computation of physical quantities. 

This paper is organized as follows. In Sec. \ref{sec:model}, we introduce the 2D ASEP-LK. The model is a continuous-time Markov process, and its time evolution is described by the master equation.
In Sec. \ref{sec:general_exp}, we focus on the steady state, which is the eigenstate of the Markov matrix with zero eigenvalue. Here, we derive the general expression for the steady state of the ASEP-LK.
In Sec. \ref{sec:torus}, we construct the exact steady state of the ASEP-LK in the periodic boundary conditions and investigate physical quantities. In Sec. \ref{sec:closed}, we obtain the exact expression of the steady state in infinitesimal Langmuir kinetics under the closed boundary conditions. Through the exact analysis, we reveal the behavior of the density distribution in response to the attachment and detachment ratio.
In Sec. \ref{sec:multi-lane}, we present the exact expression of the steady state in infinitesimal Langmuir kinetics under the multi-lane boundary conditions. Through the exact analysis, we elucidate the effect of the two-dimensionality on the quasi-one-dimensional flow. 
In Sec. \ref{sec:conclusion}, we conclude our results.

\section{Model}
\label{sec:model}

\subsection{ASEP}
The ASEP is a continuous-time Markov process in which hardcore particles hop asymmetrically in a lattice \cite{Derrida_1993,derrida1998exactly,blythe2007nonequilibrium,Crampe_2014,essler1996representations,golinelli2006asymmetric,gwa1992bethe,kim1995bethe,Golinelli_2004,Golinelli_2005,PhysRevE.85.042105,Motegi_2012,prolhac2013spectrum,prolhac2014spectrum,prolhac2016extrapolation,prolhac2017perturbative,ishiguro2023,de2005bethe,deGier_2006,deGier_2008,deGier_2011,Wen_2015,Crampe_2015,Sandow_1994,schutz1997duality,relation,schadschneider2000statistical,schadschneider2010stochastic,macdonald1968kinetics,klumpp2003traffic,bertini1997stochastic,PhysRevLett.104.230602}. In this paper, we focus on a 2D lattice with a system size $L_{\text{T}}=L_x \times L_y$ (Fig. \ref{fig:ASEP-LK_fig}). 
The updating rule is given as follows.
Particles hop to the forward site in the $x$-direction ($y$-direction) with a hopping rate $p_x$ ($p_y$) and to the backward site with a hopping rate $q_x$ ($q_y$). 
Due to the hardcore interactions, each site contains at most one particle.
We consider the combination of two types of boundary conditions: closed boundary conditions and periodic boundary conditions. 
When we consider the closed boundary conditions in the $x$-direction, a particle at site $(x,y)=(L_x,y)$ ($(x,y)=(1,y)$) cannot hop to the forward (backward) direction.
In contrast, when we consider the periodic boundary conditions, a particle at site $(x,y)=(L_x,y)$ ($(x,y)=(1,y)$) can hop to the site $(x,y)=(1,y)$ ($(x,y)=(L_x,y)$) with the rate $p_x$ ($q_x$). In this paper, we consider the three patterns of the combination of boundary conditions: torus, closed, and multi-lane boundary conditions. In the torus (closed) boundary conditions, we consider the periodic (closed) boundary conditions both in the $x$- and $y$-direction. In contrast, in the multi-lane boundary conditions, we consider the closed boundary conditions in the $x$-direction and the periodic boundary conditions in the $y$-direction.
In these boundary conditions, the number of particles $N$ is conserved throughout the time evolution. 

The state of a site $\bm{r}=(x,y)$ is represented by a Boolean number $n_{\bm{r}}$, which is set to $n_{\bm{r}} = 0$ ($n_{\bm{r}} = 1$) when the site is empty (occupied).
A configuration of the ASEP, which is denoted by $n$, is discribed by a seriese of the Boolean numbers $n=(n_{(1,1)},n_{(2,1)},\cdots,n_{(L_x,L_y)})$.
We denote the probability of the system being in a configuration $n$ at time $t$ as $P(n,t)$. The time evolution of $P(n,t)$ is described by the following master equation:
\begin{align}
    \frac{d}{dt} P(n,t) =\sum_{n'\neq n} \left[ P(n',t) W_{\text{h}}(n' \to n)-P(n,t) W_{\text{h}}(n \to n') \right],
\label{eq:mastereq}
\end{align}
where $W_{\text{h}}(n \to n')$ denotes a transition rate from $n$ to $n'$ through a hopping process.

It is also useful to rewrite the master equation (\ref{eq:mastereq}) in vector form. The state of a site $\bm{r}$ is represented by a two-dimensional vector $|n_{\bm{r}}\ket$, which equals to $|0\ket$ ($|1\ket$) when the site is empty (occupied). The $L_{\text{T}}$-fold tensor product $|n\ket=\bigotimes_{x=1}^{L_x}\bigotimes_{y=1}^{L_y}|n_{\bm{r}}\ket$ forms an orthonormal basis of the configuration space under normalization. A stochastic state vector is given by
\begin{align}
    |P(t)\ket = \sum_{n} P(n,t) |n\ket.
\end{align}
 In the Markov process, the expectation value of a physical quantity $\hat{A}$, which takes a value $A(n)$ in a configuration $n$, in a state $|P(t)\ket$ is given by
\begin{align}
\begin{split}
    \bra \hat{A} \ket &= \bra \mathcal{P} | \hat{A} | P(t) \ket \\
    &=\sum_{n}P(n,t)A(n)
    \end{split}
    \label{eq:expectation}
\end{align}
where $\bra\mathcal{P}|$ is the projection state defined as
\begin{align}
    \bra \mathcal{P} |:= \sum_{n} \bra n |.
    \label{ex:prod_st}
\end{align}
The time evolution of $|P(t)\ket$ is described by the master equation
\begin{align}
    \frac{d}{dt}|P(t)\ket = \HASEP |P(t)\ket,
    \label{eq:master_vector}
\end{align}
where the Markov matrix $\HASEP$ is given by
\begin{align}
\begin{split}
    \HASEP 
    &=\sum_{i\in\{x,y\}}\sum_{x,y} \left[ p_i \left\{ \so_{\bm{r}}^{+}\so_{\bm{r}+\bm{e}_i}^{-}- \nm_{\bm{r}}\left(1-\nm_{\bm{r}+\bm{e}_i}\right) \right\} + q_i \left\{ \so_{\bm{r}}^{-}\so_{\bm{r}+\bm{e}_i}^{+}- \left(1-\nm_{\bm{r}}\right) \nm_{\bm{r}+\bm{e}_i}\right\} \right]\\
    &=\sum_{i\in\{x,y\}}\sum_{x,y} \mathcal{M}_{\bm{r},\bm{r}+\bm{e}_i}.
\end{split}
\label{eq:ASEP}
\end{align}
Here, the subscripts of operators indicate the sites where the operators act nontrivially, and $\bm{e}_i$ represents the unit vector in $i$-direction. We introduce the half of the Pauli matrices $\so_{\bm{r}}^{x,y,z}$, the ladder operators $\so_{\bm{r}}^{\pm}=\so_{\bm{r}}^{x}\pm \im \so_{\bm{r}}^{y}$, the number operators $\nm_{\bm{r}}=1/2-\so_{\bm{r}}^z$, and the local Markov matrix 
\begin{align}
\mathcal{M}_{\bm{r},\bm{r}+\bm{e}_i}=
\begin{pmatrix}
0 & 0 & 0 & 0 \\
0 & -q_i & p_i & 0 \\
0 & q_i & -p_i & 0 \\
0 & 0 & 0 & 0 \\
\end{pmatrix}_{\bm{r},\bm{r}+\bm{e}_i}.
\end{align}
The range of the sum $\sum_{x,y}$ depends on the boundary conditions:
\begin{align}
    \begin{split}
        \sum_{x,y}=
\begin{dcases}
\sum_{x=1}^{L_x}\sum_{y=1}^{L_y} & \text{for torus boundary conditions}\\
\sum_{x=1}^{L_x-1}\sum_{y=1}^{L_y-1} & \text{for closed boundary conditions}\\
\sum_{x=1}^{L_x-1}\sum_{y=1}^{L_y} & \text{for multi-lane boundary conditions}.
\end{dcases}
\end{split}
\label{eq:sum_range}
\end{align}
The Markov matrix of the ASEP (\ref{eq:ASEP}) is regarded as the non-Hermitian Hamiltonian of the quantum spin chain. When the hopping rates are symmetric ($p_i=q_i$), the Markov matrix is equivalent to the Hamiltonian of the spin-$1/2$ Heisenberg model.

\subsection{ASEP-LK}
In this paper, we focus on the 2D ASEP with Langmuir kinetics (ASEP-LK). Namely, we introduce the attachment and detachment of particles in the bulk of the ASEP in a 2D lattice with the system size $L_{\text{T}}=L_x \times L_y$.
The updating rule is defined as follows. 
Particles hop to the forward site in the $x$-direction ($y$-direction) with a hopping rate $p_x$ ($p_x$) and to the backward site with a hopping rate $q_x$ ($q_y$). Because of the hardcore interaction, particles cannot hop to an occupied site. In the bulk, particles attach (detach) a site with a rate $\oa$ ($\od$). The schematic drawing of the ASEP-LK is shown in Fig. \ref{fig:ASEP-LK_fig}.

As in the case of the ASEP, the time evolution of the ASEP-LK is described by the master equation
\begin{align}
    \frac{d}{dt}|P(t)\ket = \HLK |P(t)\ket.
    \label{eq:master_vector_LK}
\end{align}
The Markov matrix is given by 
\begin{align}
    \HLK=\sum_{i\in\{x,y\}}\sum_{x,y} \mathcal{M}_{\bm{r},\bm{r}+\bm{e}_i} + \sum_{\bm{r}}h_{\bm{r}},
    \label{eq:Hamiltoninan_LK}
\end{align}
where $h_{\bm{r}}$ is the term that describes the attachment and detachment of particles at the site $\bm{r}$, and it is expressed as an off-diagonal magnetic field as follows:
\begin{align}
\begin{split}
    h_{\bm{r}} &= \oa \left[ \so_{\bm{r}}^{-}-\left(1-\nm_{\bm{r}}\right)\right] +\od \left[ \so_{\bm{r}}^{+}-\nm_{\bm{r}} \right] \\
    &= \begin{pmatrix}
-\oa & \od \\
\oa & -\od \\
\end{pmatrix}_{\bm{r}}.
\end{split}
\end{align}
The first term of the Markov matrix (\ref{eq:Hamiltoninan_LK}) describes the asymmetric diffusion, and the second term describes the Langmuir kinetics. 
The range of the sum $\sum_{x,y}$ depends on the boundary conditions and follows Eq. (\ref{eq:sum_range}), and that of the sum $\sum_{\bm{r}}$ is all sites.
In contrast to the case of the ASEP, the number of the particles $N$ is not conserved in the ASEP-LK.
For the subsequent discussion, we introduce the following constants
\begin{align}
    \omega:=\oa +\od, \qquad \alpha:=\frac{\oa}{\od},
\end{align}
which describe the strength of Langmuir kinetics and the attachment and detachment rate ratio, respectively.

\section{General expression of the steady state}
\label{sec:general_exp}

\subsection{Steady state of the ASEP without Langmuir kinetics}
In the case of the standard ASEP without Langmuir kinetics, the number of particles $N$ is conserved. The Markov matrix (\ref{eq:ASEP}) can be block diagonalized, and steady states exist for each subspace corresponding to the particle number $N$. 
A configuration in the $N$ particles' subspace, which is denoted by $n_N$, is represented by the positions of $N$ particles $(\bm{r_1},\bm{r_2},\cdots,\bm{r_N})$. 
We denote the basis of the $\binom{L_{\text{T}}}{N}$-dimensional subspace as $|n_N\ket=|(\bm{r_1},\bm{r_2},\cdots,\bm{r_N})\ket$.
A stochastic state vector with $N$ particles in the ASEP is given by 
\begin{align}
    |P_N(t)\ket = \sum_{n_N} P(n_N,t) |n_N\ket, \qquad \sum_{n_N} P(n_N,t)=1.
\end{align}
where $P(n_N,t)$ is the probability of the system being in $n_N$ at time $t$.

When the ASEP reaches the steady states, the master equation (\ref{eq:master_vector}) in the $N$ particles' subspace satisfies
\begin{align}
    \frac{d}{dt}|P_N(t)\ket = \HASEP|P_N(t)\ket=0.
\end{align}
Namely, the steady state is the eigenstate of the Markov matrix with a zero eigenvalue.
We denote the steady states of the ASEP with $N$ particles as
\begin{align}
    |S_N\ket = \frac{1}{Z_N}\sum_{n_N} P_{\text{st}}(n_N) |n_N\ket, \qquad Z_N := \sum_{n_N} P_{\text{st}}(n_N).
    \label{eq:st_ASEP_without_LK}
\end{align}
The steady state of the multi-dimensional ASEP is constructed in Ref. \cite{ishiguro2024exact}. In the 2D case, the steady state is given by
\begin{align}
P_{\text{st}}(n_N)=
   \begin{dcases}
        1 & \quad \text{for torus boundary conditions} \\
        \left(\frac{p_x}{q_x}\right)^{\sum_{j=1}^N x_{j}} \left(\frac{p_y}{q_y}\right)^{\sum_{j=1}^N y_{j}} & \quad \text{for closed boundary conditions}\\
        \left(\frac{p_x}{q_x}\right)^{\sum_{j=1}^N x_{j}} & \quad \text{for multi-lane boundary conditions},
    \end{dcases}
    \label{eq:lhscases}
\end{align}
where $\bm{r}_j=(x_j,y_j)$ denotes the position of the $j$-th particle ($j=1,2,\cdots,N$).

\subsection{Steady state of the ASEP-LK}
In contrast, in the case of the ASEP-LK, the number of particles $N$ is not conserved. 
Therefore, unlike the standard ASEP case, we have to consider all configuration space, and there is a unique steady state. A stochastic state vector for all configuration space is expressed as 
\begin{align}
    |P(t)\ket = \sum_{N=0}^{L_{\text{T}}} \sum_{n_N} P(n_N,t) |n_N\ket, \qquad \sum_{N=0}^{L_{\text{T}}} \sum_{n_N} P(n_N,t)=1.
\end{align}
We denote the steady state of the ASEP-LK as
\begin{align}
    |S_{\text{LK}}\ket = \frac{1}{\varXi} \sum_{N=0}^{L_{\text{T}}} \sum_{n_N} P_{\text{LK}}(n_N) |n_N\ket, \qquad \varXi:=\sum_{N=0}^{L_{\text{T}}} \sum_{n_N} P_{\text{LK}}(n_N).
\end{align}
The steady state satisfies
\begin{align}
    \HLK|S_{\text{LK}}\ket =0,
    \label{eq:master_LK}
\end{align}
which indicates that $|S_{\text{LK}}\ket$ is the zero energy eigenstate of the Markov matrix (\ref{eq:Hamiltoninan_LK}).

In the steady state of the ASEP-LK, the sum of the weight with fixed $N$ satisfies the following relation, which is proved in Sec. \ref{sec:proof}:
\begin{align}
\begin{split}
    \sum_{n_N} P_{\text{LK}}(n_N)&=\binom{L_{\text{T}}}{N} \alpha^N \\
    &=:P_{\tot}(N).
\end{split}
\label{eq:weight_formula}
\end{align}
This relation indicates that the stochastic state vector of the steady state can be written in the form 
\begin{align}
    |S_{LK}\ket = \frac{1}{\varXi} \sum_{N=0}^{L_{\text{T}}} P_{\text{T}}(N) |P_N \ket, \qquad \varXi = (1+\alpha)^{L_{\text{T}}},
    \label{eq:general_expression}
\end{align}
where $|P_N\ket$ is a stochastic state vector in the subspace with $N$ particles 
\begin{align}
    |P_N \ket = \sum_{n_N} P(n_N) |n_N\ket, \qquad \sum_{n_N} P(n_N) =1.
\end{align}
From this, we can calcurate the average density of the ASEP-LK in the steady state, which is denoted as $\rho_{\text{st}}$.
We introduce the total number operator $\hat{N}_{\text{T}}:=\sum_{\bm{r}} \hat{n}_{\bm{r}}$ and define the average density as $\rho(t)= \frac{1}{L_{\text{T}}}\bra \mathcal{P}|\hat{N}_{\text{T}}|P(t)\ket$.
From Eq. (\ref{eq:general_expression}),
\begin{align}
\begin{split}
    \rho_{\text{st}}&=\frac{1}{L_{\text{T}}} \bra \mathcal{P} | \hat{N}_{\text{T}} |S_{LK}\ket \\
    &=\frac{1}{L_{\tot}(1+\alpha)^{L_\tot}} \sum_{N=0}^{L_\tot} N P_{\tot}(N) \\
    &= \frac{\alpha}{1+\alpha},
\end{split}
\label{eq:ave_dens}
\end{align}
where we use 
\begin{align}
    \sum_{N=0}^{L_\tot} N \alpha^{N} \binom{L_{\tot}}{N} = L_{\tot} \alpha (1+\alpha)^{L_\tot-1}.
\end{align}
Eq. (\ref{eq:ave_dens}) holds for all of the torus, closed, and multi-lane boundary conditions. 

\begin{figure}[bt]
    \centering
    \includegraphics[height=6.5cm]{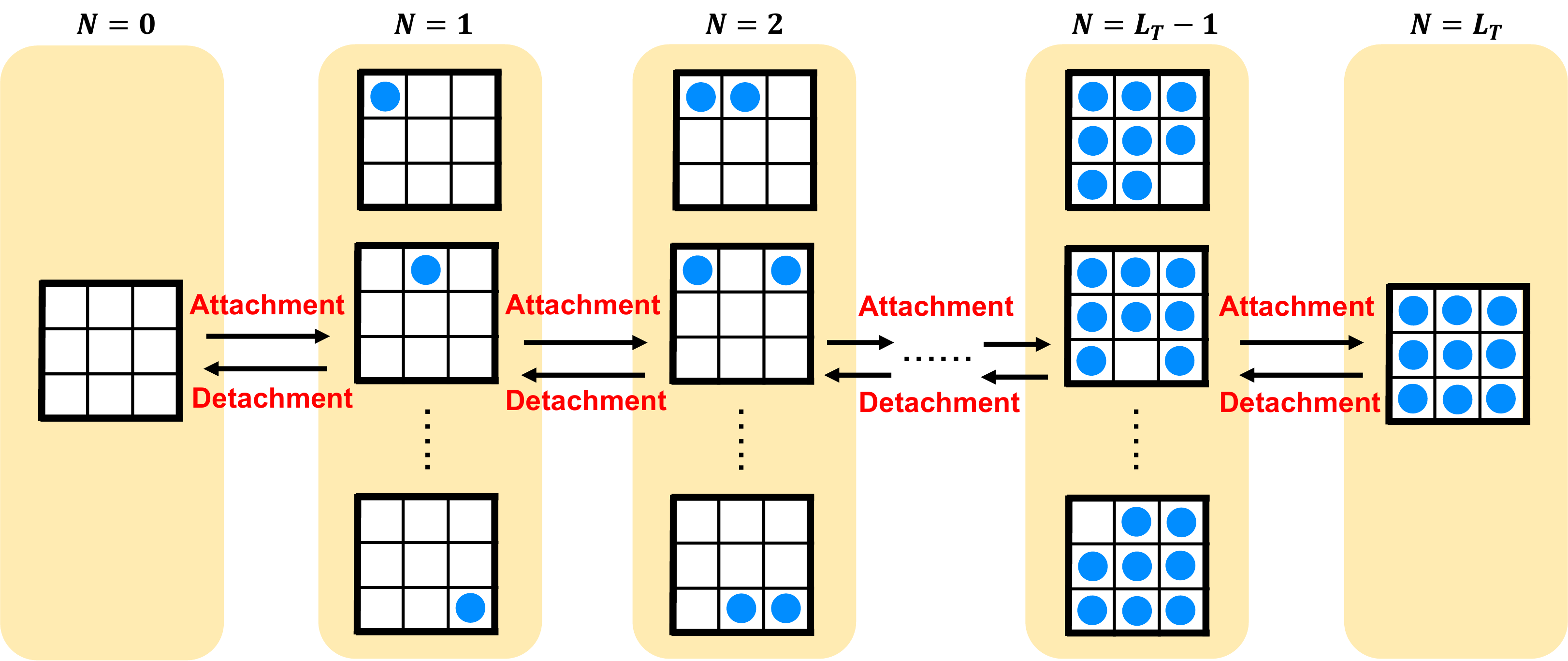}
    \caption{Transition processes of the ASEP-LK. When Langmuir kinetics rate $\omega$ is infinitesimally small, it is expected that a large number of transitions due to hopping occur between processes of the attachment and detachment of particles, achieving the steady state of the standard ASEP without Langmuir kinetics.}
    \label{fig:transition}
\end{figure}

We call Eq. (\ref{eq:general_expression}) general expression for the steady state of the ASEP-LK.
Generally, $|P_N\ket$ in Eq. (\ref{eq:general_expression}) depends on the attachment and detachment rates ($\oa,$  $\od$). 
However, under the specific conditions, $|P_N\ket$ is equal to $|S_N\ket$, where $|S_N\ket$ is the steady state of the standard ASEP without the Langmuir kinetics. That is,
\begin{align}
        |S_{LK}\ket = \frac{1}{(1+\alpha)^{L_{\text{T}}}} \sum_{N=0}^{L_{\text{T}}} P_{\text{T}}(N) |S_N \ket
        \label{eq:steady_formula}
\end{align}
provides the steady states of the ASEP-LK.
In particular, in the infinitesimal Langmuir kinetics limit ($\omega \to 0$) with the fixed attachment and detachment ratio $\alpha$, the steady state is expected to be expressed as Eq. (\ref{eq:steady_formula}). In Sec. \ref{sec:closed} and \ref{sec:multi-lane}, we numerically confirm this conjecture under the closed and multi-lane conditions. 
This is interpreted as follows based on physical intuition. In the ASEP-LK, there are two types of transition processes: hopping and Langmuir kinetics (Fig. \ref{fig:transition}).
When Langmuir kinetics rate $\omega$ is infinitesimally small, the frequency of transitions due to the hopping becomes significantly higher than that due to the attachment and detachment of particles. Therefore, it is expected that an enormous number of transitions due to hopping occur between processes of the attachment and detachment of particles, achieving the steady state of the standard ASEP without Langmuir kinetics. Thus, in the steady state of the ASEP-LK in infinitesimal Langmuir kinetics, $|S_N\ket$ is realized for each particle number sector and superposed with the weight (\ref{eq:weight_formula}).
In addition, in the torus boundary conditions, we can prove that the steady state is expressed as Eq. (\ref{eq:steady_formula}) for any value of $\omega$, as we show in Sec. \ref{sec:torus}.

\subsection{Proof of the relation (\ref{eq:weight_formula})}
\label{sec:proof}
In the following, we prove the relation (\ref{eq:weight_formula}). 
We consider a configuration with $N$ particles, which is expressed as $n_N$.
We define a set of configurations with $N+1$ particles, which is denoted as $A(n_N)$, by all configurations with $N+1$ particles that can be transitioned from $n_N$ through the attachment process.
Similarly, we also define a set of configurations with $N-1$ particles, which is denoted as $D(n_N)$, by all configurations with $N-1$ particles that can be transitioned from $n_N$ through the detachment process.

The master equation of the ASEP-LK (\ref{eq:master_LK}) can be rewritten as follows:
\begin{align}
    \begin{split}
        \frac{d P(n_N,t)}{dt} &= \sum_{n'_N \neq n_N} \left[ P(n'_N,t) W_{\text{h}}(n'_N \to n_N) - P(n_N,t) W_{\text{h}}(n_N \to n'_N) \right] \\
        & +\oa\left[\sum_{n_{N-1} \in D(n_N)} P(n_{N-1},t) - (L_{\text{T}}-N) P(n_N,t) \right] \\
        & + \od \left[\sum_{n_{N+1} \in A(n_N)} P(n_{N+1},t) -  N P(n_N,t) \right].
    \end{split}
    \label{eq:rewrited_master_LK}
\end{align}
In the right-hand side of Eq. (\ref{eq:rewrited_master_LK}), the first term describes the hopping process, the second describes the attachment process, and the third describes the detachment process. 
When we consider the steady state, the master equation is given by
\begin{align}
    \begin{split}
        \sum_{n'_N \neq n_N} &\left[ P_{\text{LK}}(n'_N) W_{\text{h}}(n'_N \to n_N) - P_{\text{LK}}(n_N) W_{\text{h}}(n_N \to n'_N) \right] \\
        & +\oa \left[ \sum_{n_{N-1} \in D(n_N)}  P_{\text{LK}}(n_{N-1}) - (L_{\text{T}}-N) P_{\text{LK}}(n_N) \right] \\
        & +\od \left[\sum_{n_{N+1} \in A(n_N)}  P_{\text{LK}}(n_{N+1}) - N P_{\text{LK}}(n_N) \right] = 0.
    \end{split}
    \label{eq:st_master_LK}
\end{align}
We take the sum of the master equations Eq. (\ref{eq:st_master_LK}) over all configurations with $N$ particles. Since the sum of the first term becomes zero:
\begin{align}
    \sum_{n_N}\sum_{n'_N \neq n_N} &\left[ P_{\text{LK}}(n'_N) W_{\text{h}}(n'_N \to n_N) - P_{\text{LK}}(n_N) W_{\text{h}}(n_N \to n'_N) \right]=0,
\end{align}
we obtain
\begin{align}
    \begin{split}
        &\oa \sum_{n_N}\left[ \sum_{n_{N-1} \in D(n_N)} P_{\text{LK}}(n_{N-1}) - (L_{\text{T}}-N) P_{\text{LK}}(n_N) \right]
        \\
        & \hspace{50pt} +\od \sum_{n_N}\left[\sum_{n_{N+1} \in A(n_N)}  P_{\text{LK}}(n_{N+1}) - N P_{\text{LK}}(n_N) \right] = 0.
    \end{split}
    \label{eq:st_master_LK_sum}
\end{align}
Here, the following is satisfied
\begin{align}
\begin{split}
    &\sum_{n_N}\sum_{n_{N-1} \in D(n_N)} P_{\text{LK}}(n_{N-1}) = (L_{\text{T}}-N+1) \sum_{n_{N-1}} P_{\text{LK}}(n_{N-1}),\\
    &\sum_{n_N}\sum_{n_{N+1} \in A(n_N)} P_{\text{LK}}(n_{N+1}) = (N+1) \sum_{n_{N+1}} P_{\text{LK}}(n_{N+1}).
    \end{split}
    \label{eq:sum_formula}
\end{align}
For a configuration $n_{N-1}$ ($n_{N+1}$), there are $L_\tot-N+1$ ($N+1$) ways to make a configuration $n_{N}$ by detaching (attaching) a particle, which corresponds to the coefficient in the right-hand side of Eq. (\ref{eq:sum_formula}).
Then, we obtain the difference equation for $P_{\text{t}}(N)$ from Eq. (\ref{eq:st_master_LK_sum}) 
\begin{align}
    (L_{\text{T}}-N+1) \alpha P_{\text{t}}(N-1) - \left[ (L_{\text{T}}-N) \alpha +N \right] P_{\text{t}}(N) + (N+1) P_{\text{t}}(N+1)=0.
    \label{eq:master_rec}
\end{align}
By substituting Eq. (\ref{eq:weight_formula}) for the left-hand side of Eq. (\ref{eq:master_rec}), we obtain
\begin{align}
\begin{split}
    \text{LHS}&=\alpha^N \left\{ (L_{\text{T}}-N+1) \binom{L_{\text{T}}}{N-1} \right. \\
    & \qquad \left. - \left[ (L_{\text{T}}-N) \alpha +N \right] \binom{L_{\text{T}}}{N} + (N+1)\alpha\binom{L_{\text{T}}}{N+1} \right\}\\
    &=0.
\end{split}
\end{align}
Namely, Eq. (\ref{eq:weight_formula}) is a solution of the difference equation (\ref{eq:master_rec}).
Therefore, in the steady state of the ASEP-LK, the relation (\ref{eq:weight_formula}) is satisfied.

\section{Steady state in the torus boundary conditions}
\label{sec:torus}

In this section, we construct the exact steady state of the ASEP-LK in the torus boundary conditions and show that it corresponds to Eq. (\ref{eq:steady_formula}). 
When considering the periodic boundary conditions, even if the hopping rates are extended inhomogeneous for each lane ($p_x \to p_x(y)$, $p_y \to p_y(x)$), we can still discuss the construction of the steady state similarly.
Therefore, we consider the Markov matrix extended as follows:
\begin{align}
    \HLK=\sum_{x=1}^{L_x}\sum_{y=1}^{L_y}\left[ \mathcal{M}_{\bm{r},\bm{r}+\bm{e}_x}(y)+\mathcal{M}_{\bm{r},\bm{r}+\bm{e}_y}(x)\right] + \sum_{\bm{r}}h_{\bm{r}},
    \label{eq:Hamiltoninan_LK_torus}
\end{align}
where
\begin{align}
\mathcal{M}_{\bm{r},\bm{r}+\bm{e}_i}(j)=
\begin{pmatrix}
0 & 0 & 0 & 0 \\
0 & -q_i(j) & p_i(j) & 0 \\
0 & q_i(j) & -p_i(j) & 0 \\
0 & 0 & 0 & 0 \\
\end{pmatrix}_{\bm{r},\bm{r}+\bm{e}_i}.
\end{align}

Here, we define the $2^{L{\text{t}}} \times 2^{L{\text{t}}}$ matrix $U$, which is originally introduced in the 1D case \cite{PhysRevE.93.042113}, as 
\begin{align}
    U=
    \underbrace{
\begin{pmatrix}
1 & 1 \\
\alpha & -1 \\
\end{pmatrix}
\otimes 
\begin{pmatrix}
1 & 1 \\
\alpha & -1 \\
\end{pmatrix}
\otimes \cdots \otimes
\begin{pmatrix}
1 & 1 \\
\alpha & -1 \\
\end{pmatrix}
}_{L_{\text{T}}},
\label{eq:matU}
\end{align}
and transform the Markov matrix (\ref{eq:Hamiltoninan_LK_torus}) as follows:
\begin{align}
\begin{split}
    \THLK & = U^{-1} \HLK U \\
    &=\sum_{x=1}^{L_x}\sum_{y=1}^{L_y}\left[ \Tilde{\mathcal{M}}_{\bm{r},\bm{r}+\bm{e}_x}(y)+\Tilde{\mathcal{M}}_{\bm{r},\bm{r}+\bm{e}_y}(x)\right] + \sum_{\bm{r}}\Tilde{h}_{\bm{r}}.
\end{split}
\label{eq:transformed_Ham}
\end{align}
$\Tilde{\mathcal{M}}_{\bm{r},\bm{r}+\bm{e}_i}(j)$ and $\Tilde{h}_{\bm{r}}$ are given by
\begin{align}
&\Tilde{\mathcal{M}}_{\bm{r},\bm{r}+\bm{e}_i}(j)= \frac{1}{1+\alpha}
\begin{pmatrix}
0 & 0 & 0 & 0 \\
-a(j) & -b(j) & c(j) & d(j) \\
a(j) & b(j) & -c(j) & -d(j) \\
0 & 0 & 0 & 0 \\
\end{pmatrix}_{\bm{r},\bm{r}+\bm{e}_i}
\label{eq:transformed_hopping}
\\
&\Tilde{h}_{\bm{r}} = - \omega \begin{pmatrix}
0 & 0 \\
0 & 1 \\
\end{pmatrix}_{\bm{r}},
\label{eq:transformed_langmuir}
\end{align}
where
\begin{align}
    \begin{split}
        &a(j)=\alpha(p_i(j)-q_i(j)), \\
        &b(j)=q_i(j)+p_i(t)\alpha, \\
        &c(j)=p_i(j)+q_i(j)\alpha, \\
        &d(j)=p_i(j)-q_i(j).
    \end{split}
\end{align}

We introduce the reference state $|\text{vac}\ket$ defined as
\begin{align}
    |\text{vac}\ket:=\underbrace{|0\ket \otimes |0\ket \otimes \cdots \otimes |0\ket}_{L_{\text{T}}},
\end{align}
where we denote an empty state as $|0\ket:=(1,0)^{T}$ and an occupied state as $|1\ket:=(0,1)^{T}$. The reference state is a zero energy eigenstate of the transformed Markov matrix (\ref{eq:transformed_Ham}):
\begin{align}
    \THLK |\text{vac}\ket = 0.
\end{align}
This is confirmed as follows. From Eq. (\ref{eq:transformed_hopping}) and (\ref{eq:transformed_langmuir}),
\begin{align}
    &\Tilde{\mathcal{M}}_{\bm{r},\bm{r}+\bm{e}_i}(j) |\text{vac}\ket = \frac{a(j)}{1+\alpha} (|\bm{r}\ket -|\bm{r}+\bm{e}_i\ket) \\
    &\Tilde{h}_{\bm{r}} |\text{vac}\ket =0,
\end{align}
where $|\bm{r}\ket$ indicates the configuration where a single particle exist at site $\bm{r}$. 
Then, 
\begin{align}
    \begin{split}
        \THLK|\text{vac}\ket &= \left\{ \sum_{x=1}^{L_x}\sum_{y=1}^{L_y}\left[ \Tilde{\mathcal{M}}_{\bm{r},\bm{r}+\bm{e}_x}(y)+\Tilde{\mathcal{M}}_{\bm{r},\bm{r}+\bm{e}_y}(x)\right] + \sum_{\bm{r}}\Tilde{h}_{\bm{r}} \right\} |\text{vac}\ket \\
        &= \sum_{y=1}^{L_y} \frac{a(y)}{1+\alpha} \sum_{x=1}^{L_x} (|\bm{r}\ket -|\bm{r}+\bm{e}_x\ket) + \sum_{x=1}^{L_x} \frac{a(x)}{1+\alpha} \sum_{y=1}^{L_y} (|\bm{r}\ket -|\bm{r}+\bm{e}_y\ket)\\
        &=0.
    \end{split}
\end{align}

By considering the inverse transformation of $|\text{vac}\ket$, we can construct the zero energy eigenstate of the Markov matrix $\HLK$:
\begin{align}
\begin{split}
    |\overline{S}_{\text{LK}}\ket &= U |\text{vac}\ket \\
    &= \underbrace{
    \begin{pmatrix}
        1 \\ \alpha
    \end{pmatrix}
    \otimes
    \begin{pmatrix}
        1 \\ \alpha
    \end{pmatrix}
    \otimes \cdots \otimes
    \begin{pmatrix}
        1 \\ \alpha
    \end{pmatrix}}_{L_{\text{T}}} \\
    &= \sum_{N=0}^{L_{\text{T}}} \alpha^N \sum_{n_N} |n_N\ket \\
    &= \sum_{N=0}^{L_{\text{T}}} \binom{L_{\text{T}}}{N} \alpha^N |S_N\ket.
\end{split}
\end{align}
Althogh $|\overline{S}_{\text{LK}}\ket$ is the eigenstates:
\begin{align}
    \HLK|\overline{S}_{\text{LK}}\ket= U\THLK |\text{vac}\ket=0,
\end{align}
$|\overline{S}_{\text{LK}}\ket$ is not a stochastic state vector and needs to be normalized. Since $\sum_{N=0}^{L_{\text{T}}} \binom{L_{\text{T}}}{N} \alpha^N = (1+\alpha)^{L_{\text{T}}}$, the steady state is given by 
\begin{align}
     |S_{\text{LK}}\ket = \frac{1}{(1+\alpha)^{L_{\text{T}}}} \sum_{N=0}^{L_{\text{T}}} \binom{L_{\text{T}}}{N} \alpha^N |S_N\ket.
     \label{eq:torus_steady_state}
\end{align}
This expression is identical to the proposed steady state (\ref{eq:steady_formula}).

In the following, we evaluate the physical quantities in the steady state.
We introduce the local density $\rho(\bm{r})$ as 
\begin{align}
    \rho(\bm{r}):=\bra\mathcal{P}|\hat{n}_{\bm{r}}|P(t)\ket,
    \label{eq:local_dens}    
    \end{align}
and the quasi-one-dimensional current $j_{x,\overline{x}}$ ($j_{y,\overline{y}}$), which represents the current in the $x$-direction ($y$-direction) at the cross-section $x=\overline{x}$ ($y=\overline{y}$), as 
\begin{align}
    j_{x,\overline{x}}:= \bra\mathcal{P}|\hat{j}_{x,\overline{x}}|P(t)\ket, \qquad j_{y,\overline{y}}:= \bra\mathcal{P}|\hat{j}_{y,\overline{y}}|P(t)\ket,
    \label{eq:quasi-current}
\end{align}
where we introduce the quadi-one-dimensional current operator $\hat{j}_{x,\overline{x}}$ ($\hat{j}_{y,\overline{y}}$) as follows:
\begin{align}
\begin{split}
    &\hat{j}_{x,\overline{x}} := \sum_{y=1}^{L_y}\left[ p_x(y) \hat{n}_{(\overline{x},y)}(1-\hat{n}_{(\overline{x}+1,y)}) - q_x(y)  (1-\hat{n}_{(\overline{x},y)})\hat{n}_{(\overline{x}+1,y)}\right], \\
    &\hat{j}_{y,\overline{y}} := \sum_{x=1}^{L_x}\left[ p_y(x) \hat{n}_{(x,\overline{y})}(1-\hat{n}_{(x,\overline{y}+1)}) - q_y(x)  (1-\hat{n}_{(x,\overline{y})})\hat{n}_{(x,\overline{y}+1)}\right].
\end{split}
\end{align}
In the case of the ASEP without Langmuir kinetics, the local density $\rho_{N}(\bm{r})$ and the quasi-one-dimensional current $j_{N;x,\overline{x}}$ ($j_{N;y,\overline{y}}$) in the steady state (\ref{eq:st_ASEP_without_LK}) is given by
\begin{align}
    &\rho_{N}(\bm{r})=\bra\mathcal{P}|\hat{n}_{\bm{r}}|S_{N}\ket =\frac{N}{L_\tot}, 
\label{eq:ASEP_without_rho} \\
\begin{split}
    &j_{N;i,\overline{i}} =\bra\mathcal{P}|\hat{j}_{i,\overline{i}}|S_{N}\ket 
     =\sum_{k=1}^{L_k} \left[ p_i(k)-q_i(k) \right] \frac{N(L_\tot-N)}{L_\tot (L_\tot-1)},  \quad (i,k=x,y,\;\; k\neq i). 
     \end{split}
\label{eq:ASEP_without_current}
\end{align}
Then, we calculate the local density for the AESP-LK from Eq. (\ref{eq:torus_steady_state}), Eq. (\ref{eq:local_dens}), and Eq. (\ref{eq:ASEP_without_rho}) as follows: 
\begin{align}
\begin{split}
    \rho_{\text{st}}(\bm{r}) &= \bra\mathcal{P}|\hat{n}_{\bm{r}}|S_{\text{LK}}\ket \\
    &= \frac{1}{(1+\alpha)^{L_{\text{T}}}} \sum_{N=0}^{L_{\text{T}}} \binom{L_{\text{T}}}{N} \alpha^N \rho_N(\bm{r}) \\
    &= \frac{\alpha}{1+\alpha}.
\end{split}
\end{align}
This is equal to the average density (\ref{eq:ave_dens}) since the stationary distribution of particles is spatially uniform in the torus boundary conditions case. 
Similarly, the quasi-one-dimensional current $j_{x,\overline{x}}$ ($j_{y,\overline{y}}$) in the steady state of the ASEP-LK is obtained from Eq. (\ref{eq:torus_steady_state}), Eq. (\ref{eq:local_dens}), and Eq. (\ref{eq:ASEP_without_current}) as follows:
\begin{align}
    \begin{split}
j_{i,\overline{i}}&:=\bra\mathcal{P}|\hat{j}_{i,\overline{i}}|S_{LK}\ket \\
        &=\frac{1}{(1+\alpha)^{L_{\text{T}}}} \sum_{N=0}^{L_{\text{T}}} \binom{L_{\text{T}}}{N} \alpha^N \hat{j}_{N;i,\overline{i}} (\bm{r})\\
        &=\sum_{k=1}^{L_k} [p_i(k)-q_i(k)] \frac{\alpha}{1+\alpha} \left( 1- \frac{\alpha}{1+\alpha}\right)\\
        &=\sum_{k=1}^{L_k} [p_i(k)-q_i(k)] \rho_{\text{st}}(1-\rho_{\text{st}}) \qquad (i,k=x,y,\;\; k\neq i). 
    \end{split}
    \label{eq:torus_current}
\end{align}
In the standard ASEP without Langmuir kinetics, the quasi-one-dimensional current (\ref{eq:ASEP_without_current}) depends on the system size $L_{\tot}$. In contrast, in the case of the ASEP-LK, the quasi-one-dimensional current (\ref{eq:torus_current}) does not depend on the system size $L_\tot$. 
This result is consistent with the 1D case \cite{Ezaki_2012_LK,PhysRevE.93.042113}.

\section{Steady state in the closed boundary conditions}
\label{sec:closed}

In this section, we consider the closed boundary conditions where the Markov matrix is given by
\begin{align}
        \HLK=\sum_{x=1}^{L_x-1}\sum_{y=1}^{L_y-1}\left[ \mathcal{M}_{\bm{r},\bm{r}+\bm{e}_x}+\mathcal{M}_{\bm{r},\bm{r}+\bm{e}_y}\right] + \sum_{\bm{r}}h_{\bm{r}}.
    \label{eq:Hamiltoninan_LK_closed}
\end{align}
\begin{figure}[bt]
    \centering
    \includegraphics[height=12.0cm]{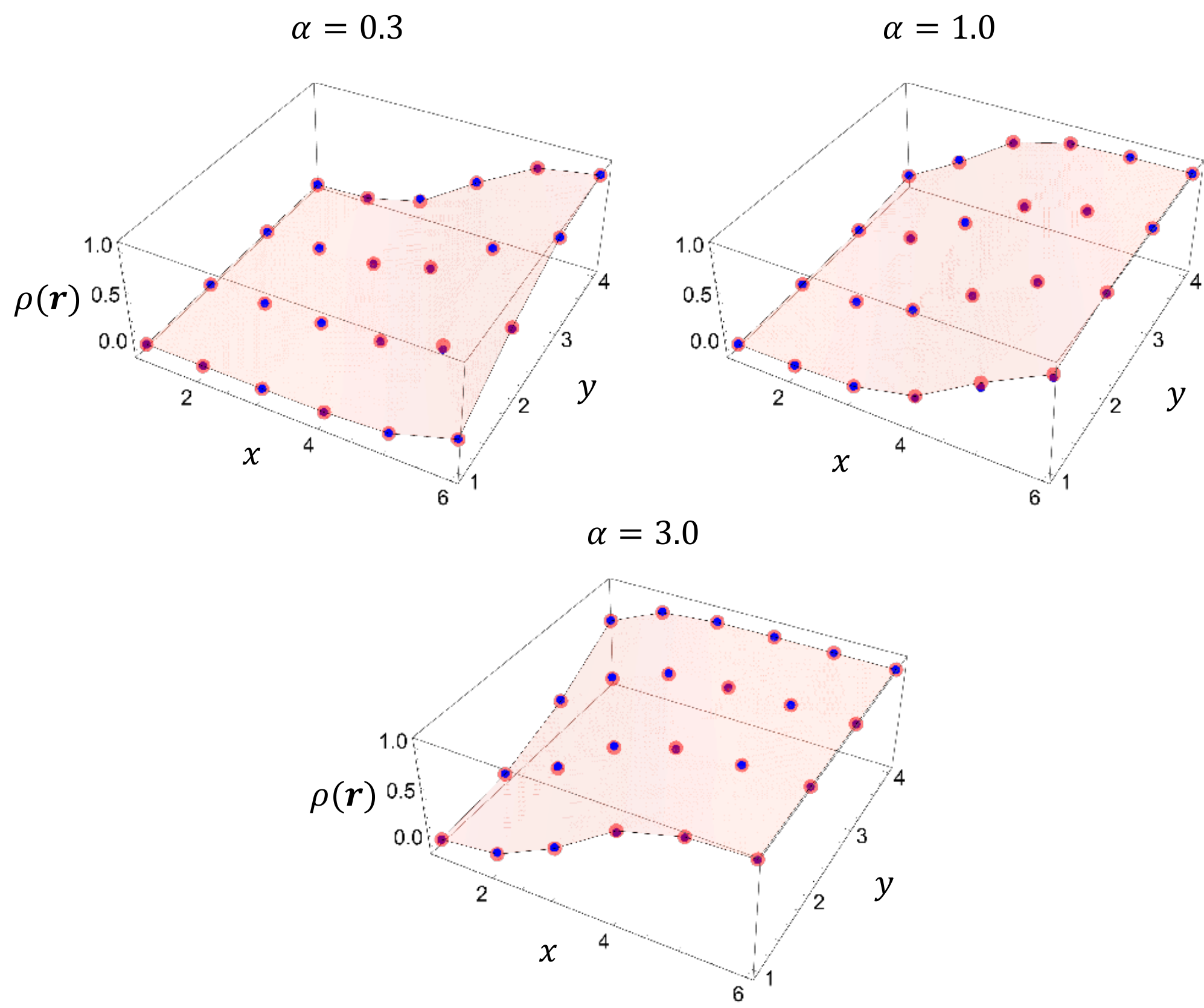}
    \caption{Local density of the ASEP-LK in the closed boundary conditions for various $\alpha$. We set the parameters as $(p_x,q_x,p_y,q_y)=(1.0,0.1,0.8,0.1)$ and $(L_x,L_y)=(6,4)$. Red dots represent the results from Eq. (\ref{eq:dens_closed}), and blue dots represent those from Monte Carlo simulations.}
    \label{fig:closed_dens_alpha}
\end{figure}
Unlike the torus boundary conditions, $|P_N\ket$ in the general expression for the steady state (\ref{eq:general_expression}) depends on $\oa$ and $\od$.
However, in the case of the infinitesimal Langmuir kinetics ($\omega\to 0$) with the fixed ratio of the attachment and detachment rates $\alpha$, Eq. (\ref{eq:steady_formula}) is expected to be the steady state.
We numerically confirmed that 
\begin{align}
        |S_{LK}\ket = \frac{1}{(1+\alpha)^{L_{\text{T}}}} \sum_{N=0}^{L_{\text{T}}}  \sum_{n_N}P_{\text{T}}(N)\left(\frac{p_x}{q_x}\right)^{\sum_{j=1}^N x_{j}} \left(\frac{p_y}{q_y}\right)^{\sum_{j=1}^N y_{j}}|n_N \ket
        \label{eq:steady_closed}
\end{align}
gives the steady state of the ASEP-LK with the closed boundary conditions.

\begin{figure}[bt]
    \centering
    \includegraphics[height=12.0cm]{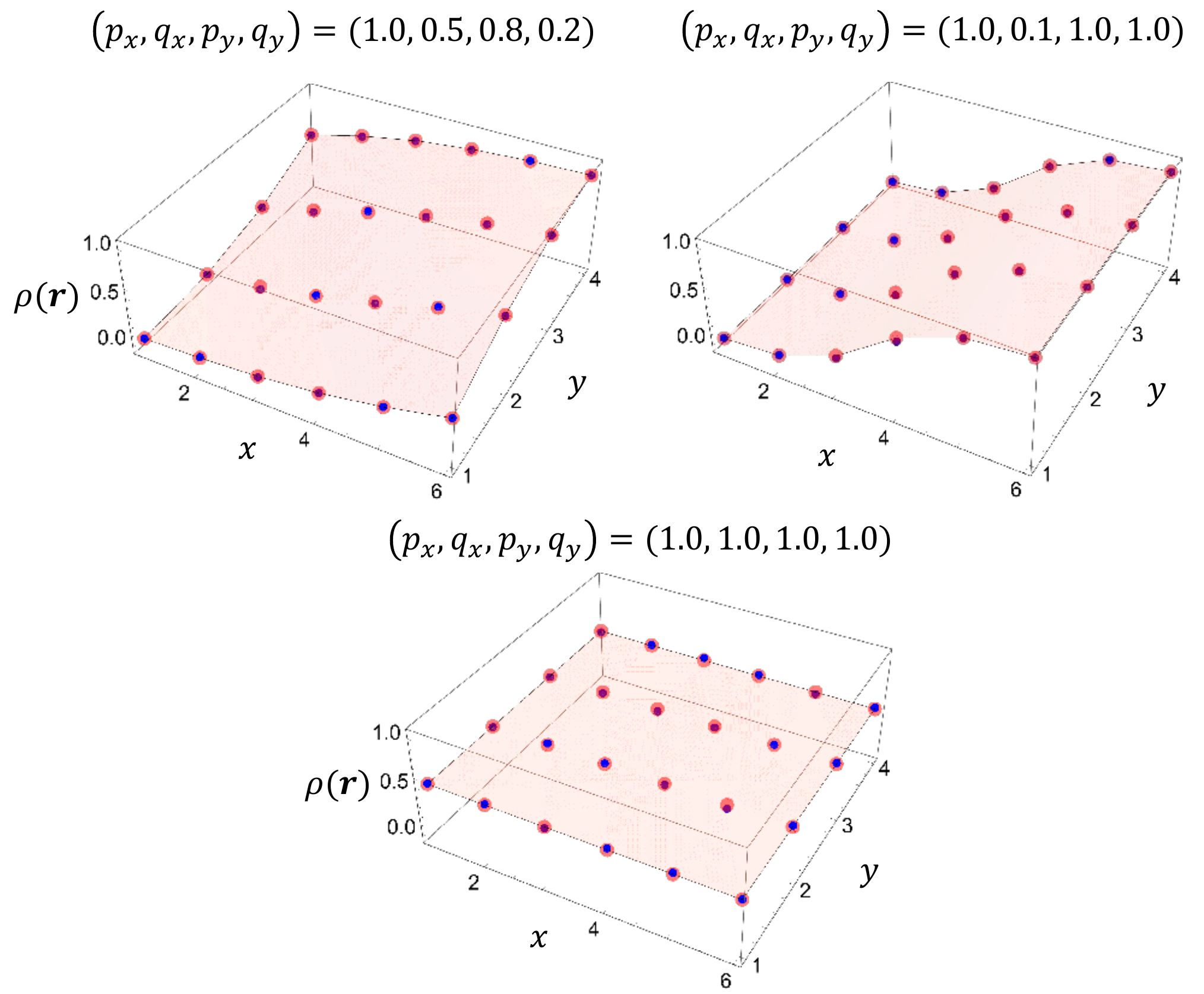}
    \caption{Local density of the ASEP-LK in the closed boundary conditions for various hopping rates $(p_x,q_x,p_y,q_y)$. We set the parameters as $\alpha=1.0$ and $(L_x,L_y)=(6,4)$. Red dots represent the results from Eq. (\ref{eq:dens_closed}), and blue dots represent those from Monte Carlo simulations.}
    \label{fig:closed_dens_hopping}
\end{figure}

In the standard ASEP without Langmuir kinetics, the local density $\rho_N(\bm{r})$ of the steady state is given by 
\begin{align}
\begin{split}
    \rho_N(\bm{r}) &= \bra\mathcal{P}|\hat{n}_{\bm{r}}|S_N\ket\\
    &=\frac{1}{Z_N} \sum_{\Tilde{n}_N} \left(\frac{p_x}{q_x}\right)^{\sum_{j=1}^N x_j} \left(\frac{p_y}{q_y}\right)^{\sum_{j=1}^N y_j},
\end{split}    
\end{align}
where $\sum_{\Tilde{n}_N}$ represents the sum over all configurations with $n_{\bm{r}}=1$ in the subspace of $N$ particles. 
Then, the local density of the steady state in the ASEP-LK is expressed as follows:
\begin{align}
\begin{split}    
    \rho_{\text{st}}(\bm{r}) &= \bra\mathcal{P}|\hat{n}_{\bm{r}}|S_{\text{LK}}\ket\\
    &=\frac{1}{(1+\alpha)^{L_{\text{T}}}} \sum_{N=0}^{L_{\text{T}}} \binom{L_{\text{T}}}{N} \alpha^N \rho_N(\bm{r}). 
    \label{eq:dens_closed}
\end{split}
\end{align}

Fig.\ref{fig:closed_dens_alpha} and Fig. \ref{fig:closed_dens_hopping} show the local density of the steady state in the ASEP-LK, which are computed from both the exact expression (\ref{eq:dens_closed}) and Monte Carlo simulations. Red dots represent the results from Eq. (\ref{eq:dens_closed}), and blue dots represent those from Monte Carlo simulations.
These figures indicate that the results obtained from Eq.  (\ref{eq:dens_closed}) are consistent with those computed from Monte Carlo simulations.
These results strongly suggest that Eq. (\ref{eq:steady_closed}) gives the exact stationary state of the ASEP-LK.

In the following, we discuss the properties of the steady state with respect to the attachment and detachment ratio $\alpha$ and hopping rates $(p_x,q_x,p_y,q_y)$.
Fig. \ref{fig:closed_dens_alpha} shows the local density for various attachment and detachment ratios $\alpha$. 
The size of the high-density and the low-density areas change depending on the value $\alpha$. As $\alpha$ increases (decreases), the size of the high-density area becomes large (small).
Fig. \ref{fig:closed_dens_hopping} shows the local density for various hopping rates $(p_x,q_x,p_y,q_y)$. The magnitude of the gradient between the high-density and low-density areas changes depending on the value of the hopping rates. As the asymmetry of the hopping rates increases, the gradient becomes steeper.

\section{Steady state in the multi-lane boundary conditions}
\label{sec:multi-lane}

\begin{figure}[tb]
    \centering
    \includegraphics[height=6.0cm]{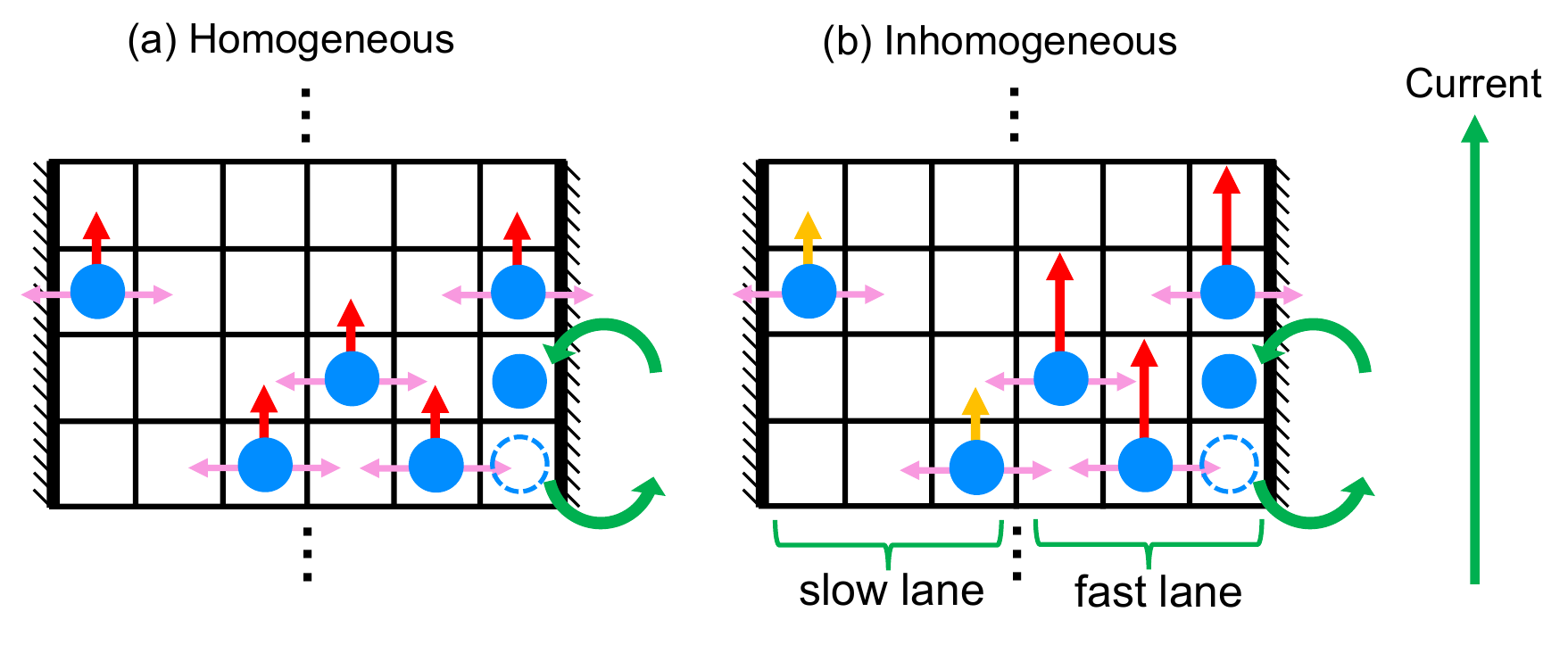}
    \caption{Examples of (a) Homogeneous and (b) inhomogeneous multi-lane ASEP with Langmuir kinetics. The hopping rate in the $y$-direction can be extended to be inhomogeneous for each lane in the multi-lane boundary conditions.}
    \label{fig:inhomogeneous}
\end{figure}
In this section, we discuss the steady state in the multi-lane boundary conditions, considering the closed boundary conditions in the $x$-direction and the periodic boundary conditions in the $y$-direction.
In the case of the multi-lane boundary condition, the steady state can be constructed even if the hopping rate in the $y$-direction is extended to be inhomogeneous for each lane ($p_y \to p_y(x)$) as shown in Fig. \ref{fig:inhomogeneous}.
The Markov matrix of the ASEP-LK in the multi-lane boundary conditions is given by
\begin{align}
        \HLK=\sum_{x=1}^{L_x-1}\sum_{y=1}^{L_y}\left[ \mathcal{M}_{\bm{r},\bm{r}+\bm{e}_x}+\mathcal{M}_{\bm{r},\bm{r}+\bm{e}_y}(x)\right] + \sum_{\bm{r}}h_{\bm{r}}.
        \label{eq:Hamiltonian_multi}
\end{align}
As in the case of the closed boundary conditions, Eq. (\ref{eq:steady_formula}) is expected to give the steady state of the ASEP-LK (\ref{eq:Hamiltonian_multi}) in the infinitesimal Langmuir kinetics $\omega \to 0$ with the fixed attachment and detachment ratio $\alpha$.
We numerically confirmed that 
\begin{align}
            |S_{LK}\ket = \frac{1}{(1+\alpha)^{L_{\text{T}}}} \sum_{N=0}^{L_{\text{T}}}  \sum_{n_N}P_{\text{T}}(N)\left(\frac{p_x}{q_x}\right)^{\sum_{j=1}^N x_{j}}|n_N \ket
        \label{eq:steady_multi}
\end{align}
is the steady state of the ASEP-LK in the multi-lane boundary conditions.

\begin{figure}[tb]
    \centering
    \includegraphics[height=12.0cm]{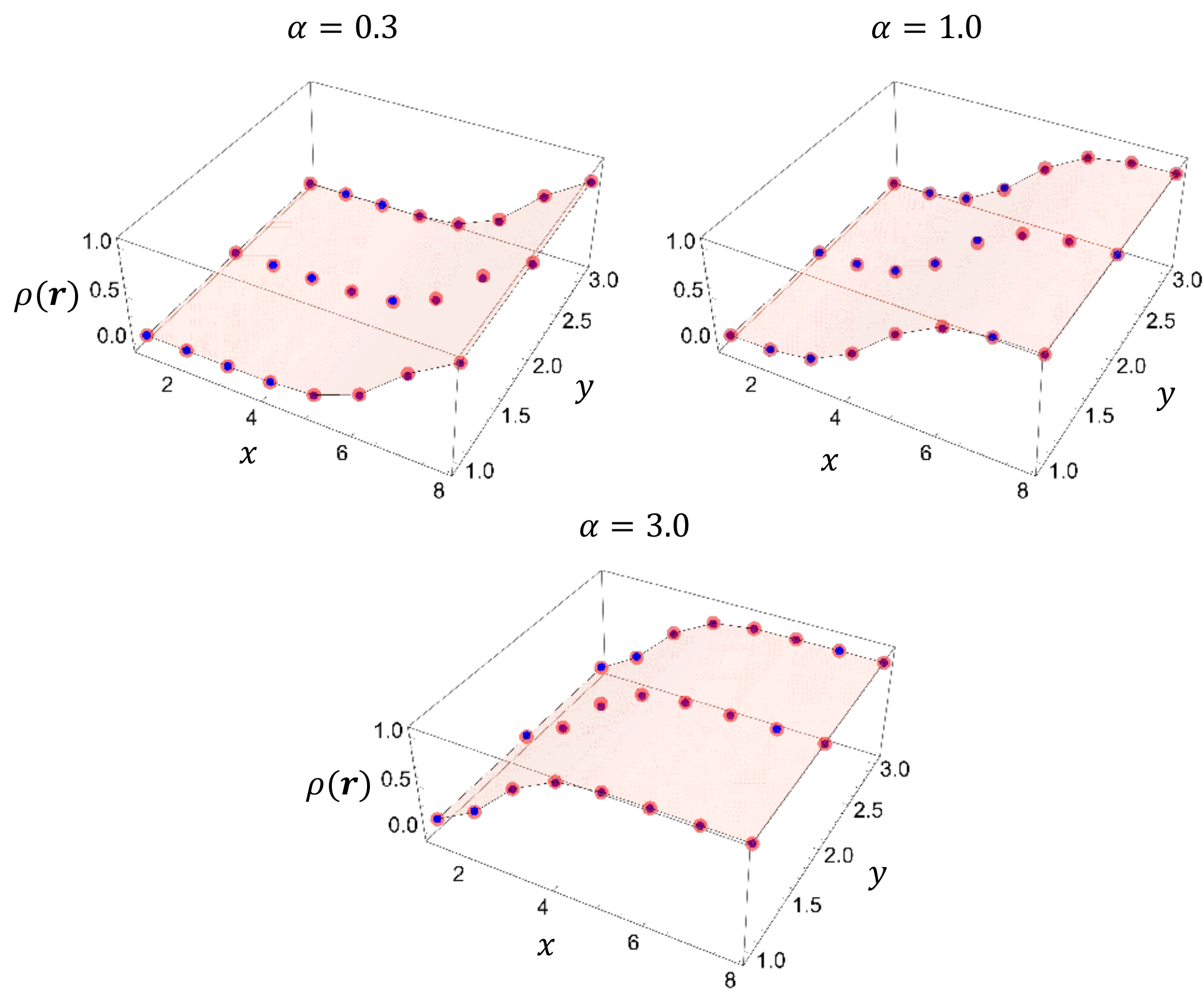}
    \caption{Local density of the ASEP-LK in the multi-lane boundary conditions for various $\alpha$. We set the parameters as $(p_x,q_x)=(1.0,0.1)$ and $(L_x,L_y)=(8,3)$. Red dots represent the results from Eq. (\ref{eq:dens_multi}), and blue dots represent those from Monte Carlo simulations.}
    \label{fig:multi_dens_alpha}
\end{figure}

\begin{figure}[tbh]
    \centering
    \includegraphics[height=12.0cm]{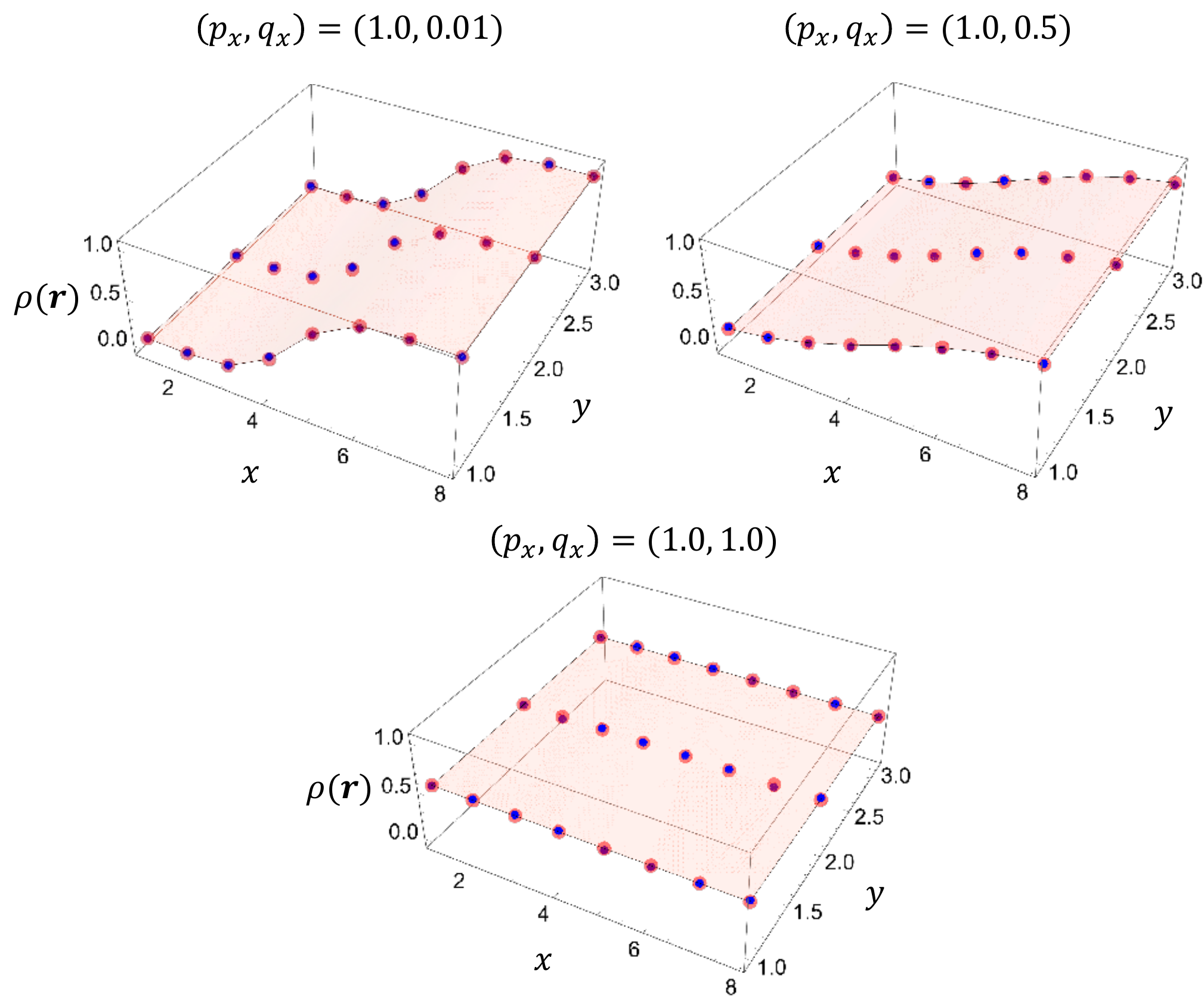}
    \caption{Local density of the ASEP-LK in the multi-lane boundary conditions for various hopping rates $(p_x,q_x)$. We set the parameters as $\alpha=1.0$ and $(L_x,L_y)=(8,3)$.  Red dots represent the results from Eq. (\ref{eq:dens_multi}), and blue dots represent those from Monte Carlo simulations.}
    \label{fig:multi_dens_hopping}
\end{figure}

In the standard ASEP without Langmuir kinetics, the local density $\rho_N(\bm{r})$ of the steady state (\ref{eq:st_ASEP_without_LK}) is given by
\begin{align}
\begin{split}    
    \rho_N(\bm{r}) &= \bra\mathcal{P}|\hat{n}_{\bm{r}}|S_N\ket\\
    &=\frac{1}{Z_N} \sum_{\Tilde{n}_N} \left(\frac{p_x}{q_x}\right)^{\sum_{j=1}^N x_j},
\end{split}
\label{eq:dens_ASEP_multi}
\end{align}
where $\sum_{\Tilde{n}_N}$ represents the sum over all configurations with $n_{\bm{r}}=1$ in the subspace of $N$ particles. 
Similary, the quasi-one-dimensional current $j_{N;y,\overline{y}}$ of the steady state (\ref{eq:st_ASEP_without_LK}) is 
\begin{align}
    \begin{split}
        j_{N;y,\overline{y}} &=\bra\mathcal{P}|\hat{j}_{y,\overline{y}}|S_{N}\ket \\
        &=\frac{1}{Z_N} \sum_{x=1}^{L_x} \left[ \sum_{\overline{n}_{N;\text{f}}} p_y(x) \left(\frac{p_x}{q_x} \right)^{\sum_{j=1}^N x_j} -\sum_{\overline{n}_{N;\text{b}}} q_y(x) \left(\frac{p_x}{q_x} \right)^{\sum_{j=1}^N x_j}\right]
    \end{split}
\label{eq:current_ASEP_multi},
\end{align}
where $\sum_{\overline{n}_{N;\text{f}}}$ ($\sum_{\overline{n}_{N;\text{b}}}$) represents the sum over all configurations with $n_{\bm{r}}=1$ and $n_{\bm{r}+\bm{e}_y}=0$ ($n_{\bm{r}}=0$ and $n_{\bm{r}+\bm{e}_y}=1$) in the subspace of $N$ particles. 
Then, the local density $\rho_{\text{st}}(\bm{r}) $ and the quasi-one-dimensional current in the $y$-direcsion $j_{y,\overline{y}}$ of the steady state in the ASEP-LK (\ref{eq:steady_multi}) is expressed as follows:
\begin{align}
\begin{split}    
    &\rho_{\text{st}}(\bm{r}) = \bra\mathcal{P}|\hat{n}_{\bm{r}}|S_{\text{LK}}\ket \\
    &\hspace{29pt} =\frac{1}{(1+\alpha)^{L_{\text{T}}}} \sum_{N=0}^{L_{\text{T}}} \binom{L_{\text{T}}}{N} \alpha^N \rho_N(\bm{r}), 
\end{split}
\label{eq:dens_multi} 
\\
\begin{split}
&j_{y,\overline{y}}=\bra\mathcal{P}|\hat{j}_{y,\overline{y}}|S_{LK}\ket 
 \\
 &\hspace{16pt} =\frac{1}{(1+\alpha)^{L_{\text{T}}}} \sum_{N=0}^{L_{\text{T}}} \binom{L_{\text{T}}}{N} \alpha^N j_{N;y,\overline{y}}.
\end{split}
\label{eq:current_multi}
\end{align}

Fig. \ref{fig:multi_dens_alpha} and Fig. \ref{fig:multi_dens_hopping} show the local density of the steady state in the multi-lane boundary conditions, which are obtained from both the exact expression (\ref{eq:dens_multi}) and Monte Carlo simulations. Red dots represent the results from Eq. (\ref{eq:dens_multi}), and blue dots represent those from Monte Carlo simulations.
These figures indicate that the results obtained from Eq.  (\ref{eq:dens_multi}) are consistent with those computed from Monte Carlo simulations. 
These results strongly suggest that Eq. (\ref{eq:steady_multi}) is the exact stationary state of the ASEP-LK in the multi-lane boundary conditions.

\begin{figure}[tb]
    \centering
    \includegraphics[height=10.0cm]{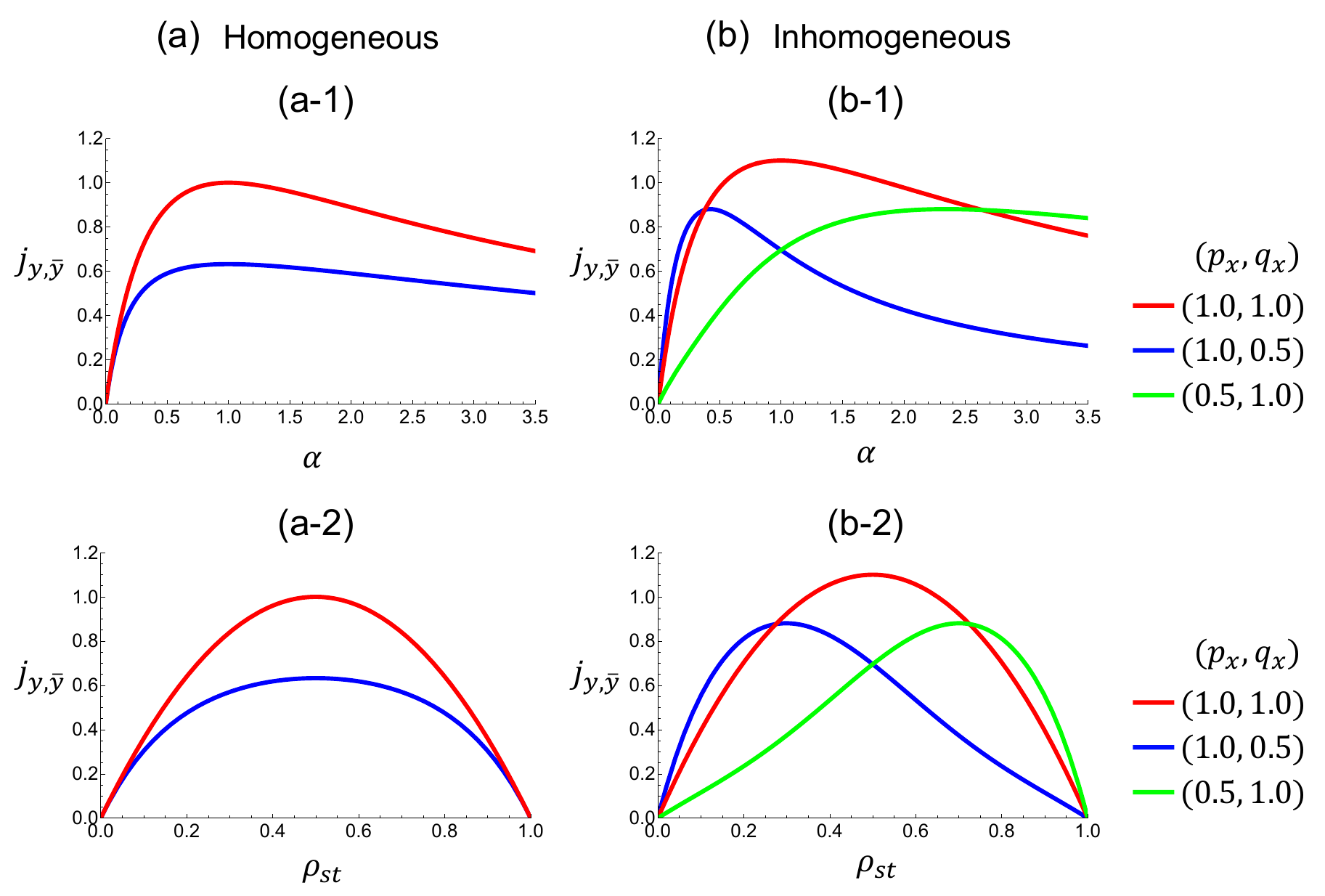}
    \caption{Quasi-one-dimensional current $j_{y,\overline{y}}$ with respect to the attachment and detachment ratio $\alpha$ and the average density $\rho_{\text{st}}$. We set the parameters as $(L_x,L_y)=(8,3)$, $q_y=0$, and $\overline{y}=1$. (a) Homogeneous hopping rate in $y$-direction ($p_y(x)=0.5$). (b) Inhomogeneous hopping rate in $y$ direction ($p_y(x)=0.1$ for $1\le x\le L_x/2$, and $p_y(x)=1.0$ for $L_x/2< x\le L_x$).}
    \label{fig:multi_current}
\end{figure}

Fig. \ref{fig:multi_dens_alpha} shows the local density for various attachment and detachment ratio $\alpha$, and Fig. \ref{fig:multi_dens_hopping} shows that for various hopping rates $(p_x,q_x)$.
As in the case of the closed boundary conditions, the size of the high- and low-density areas depends on the value $\alpha$, and the magnitude of the gradient between high- and low-density areas depends on the asymmetricity of the hopping rates $(p_x,q_x)$.

Fig. \ref{fig:multi_current} shows the quasi-one-dimensional current $j_{y,\overline{y}}$ in the steady state, which is calculated from Eq. (\ref{eq:current_multi}). Figs. \ref{fig:multi_current} (a) show the results of the homogeneous multi-lane ASEP-LK where the hopping rates are uniform ($p_y(x)=0.5$), and Figs. \ref{fig:multi_current} (b) shows those of the inhomogeneous multi-lane ASEP-LK where the hopping rates are non-uniform for each lane ($p_y(x)=0.1$ for $1\le x \le L_x/2$, and $p_y(x)=0.1$ for $L_x/2 < x \le 1$).
From the results, we find that the quasi-one-dimensional current depends on the hopping rates perpendicular to the current direction. 
The behavior differs depending on whether the hopping rate along the current direction is homogeneous or inhomogeneous across the lanes.
Figs. \ref{fig:multi_current} (a-1) and (a-2) show that, in the case of the hopping rates are homogeneous, the quasi-one-dimensional current is larger when the hopping rates in the $x$-direction are symmetric ($p_x=q_x$) compared to when they are asymmetric ($p_x \neq q_x$). 
In contrast, in the hopping rates $p_y(x)$ are inhomogeneous (Figs. \ref{fig:multi_current} (b-1) and (b-2)), there are regions where the current takes on larger value when the hopping rates are asymmetric than when they are symmetric.

\section{Conclusion}
\label{sec:conclusion}
In this paper, we considered the 2D ASEP-LK, where particle number conservation is violated. 
We obtained the general expression for the steady state of the ASEP-LK, which indicates that the steady state is constructed from the superposition of the steady states of the ASEP without Langmuir kinetics.
By applying the result of the 2D ASEP in Ref. \cite{ishiguro2024exact}, we obtained the exact steady state of the ASEP-LK under the periodic boundary conditions and that in infinitesimal Langmuir kinetics under the closed and multi-lane boundary conditions.
Based on the analytical expression, we evaluated the density and the quasi-one-dimensional current in the steady state.
In the closed and multi-lane boundary conditions, the density distribution is non-uniform. High- and low-density areas co-exist, and the size depends on the attachment and detachment ratio $\alpha$. 
In addition, we clarified the effect of two-dimensionality on the quasi-one-dimensional current. The quasi-one-dimensional current in the multi-lane boundary conditions depends on the hopping rates perpendicular to the current direction. When the hopping rates along the current are uniform among lines, the asymmetry of the perpendicular hopping rates reduces the current. However, when the hopping rates along the current are inhomogeneous among lanes, the asymmetry of the perpendicular hopping rates can become larger, depending on the value of $\alpha$.

\section*{Acknowledgement}
The authors thank Masataka Watanabe for fruitful discussions. This work was supported by JSPS KAKENHI Grant Number JP24K16976.

\section*{References}

\bibliographystyle{iopart-num-long_mod}
\bibliography{ref}

\end{document}